# Spectroscopic Evaluation of Charge-transfer Doping and Strain in Graphene/$MoS_2$ Heterostructures

*Rahul Rao,[1,2]* Ahmad E. Islam,[1,2] Simranjeet Singh,[3] Rajiv Berry,[1] Roland K Kawakami,[4] Benji Maruyama,[1] Jyoti Katoch[3]*

[1]*Materials and Manufacturing Directorate, Air Force Research Laboratory, Wright Patterson AFB, OH, 45433, USA*

[2]*UES Inc., Dayton OH, 45425, USA*

[3]*Department of Physics, Carnegie Mellon University, Pittsburgh, PA 15213, USA*

[4]*Department of Physics, The Ohio State University, Columbus, OH, 43210, USA*

## Abstract

It is important to study the van der Waals interface in emerging vertical heterostructures based on layered two-dimensional (2D) materials. Being atomically thin, 2D materials are susceptible to significant strains as well as charge transfer doping across the interfaces. Here we use Raman and photoluminescence (PL) spectroscopy to study the interface between monolayer graphene/$MoS_2$ heterostructures prepared by mechanical exfoliation and layer-by-layer transfer. By using correlation analysis between the Raman modes of graphene and $MoS_2$ we show that both layers are subjected to compressive strain and charge transfer doping following mechanical exfoliation and thermal annealing. Furthermore, we show that both strain and carrier concentration can be modulated in the heterostructures with additional thermal annealing. Our study highlights the importance of considering both mechanical and electronic coupling when characterizing the interface in van der Waals heterostructures, and demonstrates a method to tune their electromechanical properties.

## 1. Introduction

The advent of 2D layered materials beyond graphene has initiated a new field of research in vertical and lateral heterostructures wherein the stacking between the layers occurs through the weakly attractive van der Waals force. This allows the creation of a seemingly limitless number of artificial architectures where the properties of each layer can be combined towards unique applications[1] in electronics,[2] optoelectronics[3] and photovoltaics.[4] Among the layered 2D materials, graphene and the semiconducting transition metal dichalcogenide $MoS_2$ are arguably the most heavily studied. Graphene is an atomically thin semi-metal, while monolayer $MoS_2$ is a direct band gap semiconductor with an optical gap of ~1.8 eV. The interaction between graphene and $MoS_2$ in vertical heterostructures has been reported in several publications in the literature, with some results in conflict with each other. The transfer of both holes[5] and electrons[6] from $MoS_2$ to graphene has been reported, as well as reports of no charge transfer at all between the two.[7, 8] Here we show that the discrepancies between the reported results may be explained by considering strain.

Being atomically thin, 2D layers can be subjected to significant in-plane and out-of-plane strains. The lattice deformation caused by such strains could adversely affect the thermal[9] and electrical[10, 11] conductivities of the heterostructure, although studies have shown that their optoelectronic properties can be manipulated by strain engineering.[12]

Both compressive and tensile strains have been found in graphene and $MoS_2$ within a vertical heterostructure.[10, 13, 14] These strains likely originate during the mechanical transfer process, although thermal annealing and substrate interactions could impose additional strains on the 2D layers.

Raman and PL spectroscopy are arguably the most useful methods to study both strain and doping in graphene and $MoS_2$. The frequencies of the two major Raman peaks in graphene (*G* and *G'* peaks) and in $MoS_2$ (*E'* and *A'* peaks) are influenced by charge transfer doping and strain, and correlation analysis between the two modes offers a way to decouple their effects.[15-22] That is, by plotting the frequencies of the *G'* (*E'*) band against the *G* (*A')* band in graphene ($MoS_2$), one can establish the extent of strain versus doping in each layer. In addition to Raman peaks, monolayer $MoS_2$ exhibits strain- and doping-dependent photoluminescence (PL) emission, offering an additional means to study the interaction between graphene and $MoS_2$.

Here, by performing correlation analysis using Raman and PL spectroscopy, we show how strain and doping both affect graphene and $MoS_2$ when graphene is placed onto $MoS_2$ in a vertical heterostructure. We map two monolayer graphene/$MoS_2$ heterostructures and observe the following upon photoexcitation – (i) electrons are transferred from $MoS_2$ to graphene, and (ii) both graphene and $MoS_2$ are strained compressively. The degree of strain and doping varies between the two heterostructures and we show how a simple thermal annealing treatment can modulate both strain and doping. Our results highlight the effects of mechanical and electronic coupling between atomically thin layers subjected to van der Waals interactions and at the same time demonstrate a practical method to tune the electronic and mechanical properties of van der Waals heterostructures.

## 2. Sample Preparation

Bulk graphite and $MoS_2$ were mechanically exfoliated with tape to obtain monolayer graphene and $MoS_2$ on to separate $SiO_2$/Si substrates. For fabricating graphene/$MoS_2$ heterostructures a polymer-based pick-up technique was utilized for the transfer process.[23, 24] First, a small piece of polydimethylsiloxane (PDMS) stamp was attached on a glass slide. Separately, a polycarbonate (PC) film was prepared on another glass slide by putting a few drops of PC solution and spreading it across the slide. Then PC film was peeled off from the glass slide and attached onto the PDMS stamp. This PC/PDMS glass slide was then used to first pick up graphene from $SiO_2$ substrate. For this, PC/PDMS glass slide was brought in contact with the graphene and then the polymer was softened by heating it to 90 °C. Subsequently, the polymer was cooled to room temperature to pick up graphene from the $SiO_2$ substrate. Next, this PC film carrying graphene was aligned with the monolayer $MoS_2$ on $SiO_2$ substrate under an optical microscope and then PC film melted onto $MoS_2$ by heating it up to 150 °C. Finally, the PC polymer was removed by rinsing the substrate in chloroform for about 10−15 min to obtain a graphene/$MoS_2$ heterostructure.

## 3. Results

Both monolayer graphene and $MoS_2$ are prepared on Si/$SiO_2$ substrates by mechanical exfoliation, followed by transfer of the graphene on to $MoS_2$ (see details in the Sample Preparation section). After transfer, the samples first undergo annealing under vacuum (370 ºC at $10^{-5}$ torr for 3 hours) and then under a mixture of Ar and $H_2$ (350 ºC for 3 hours) to remove trapped residues and to improve the coupling between the two layers. Figure 1a shows an optical microscope image collected from a graphene/$MoS_2$ heterostructure (Het-1). The graphene layer can be identified by the slightly darker contrast compared to the underlying $SiO_2$ substrate; the dashed outline serves as a guide. Figure 1b shows a two-dimensional (2D) Raman intensity map of the first order *G* peak in graphene, corresponding to in-plane vibrations of the $sp^2$ bonded carbon atoms. All spectra are collected with 514.5 nm laser excitation in a Renishaw Raman microscope with a 100x objective lens (producing a 600 nm spot size). The laser power is kept under 70 μW to minimize sample heating. The intensity map of the *E'* peak in $MoS_2$ is shown in Figure 1c. Similar to the graphene *G* peak, the *E'* peak in $MoS_2$ corresponds to in-plane vibrations. A comparison between Figures 1b and 1c shows the presence of graphene and $MoS_2$ on $SiO_2$ outside the heterostructure (also indicated by the arrows in the optical microscope image in Figure 1a and hereafter referred to as bare graphene and $MoS_2$). This allows us to compare between their spectra on the bare substrate and in the vertical heterostructure. The other prominent peak in the Raman spectrum of graphene is the *G'* peak, which arises from scattering of photo-excited electrons by two transverse optical phonons. The *G'* peak is typically much more intense than the *G* peak in the Raman spectrum from monolayer graphene, as seen in the intensity ratio ($I_{G'}/I_G$) map in Figure 1d. $I_{G'}/I_G$ decreases by a factor of 2-3 in the heterostructure and is attributed to charge transfer doping.[25]

Both graphene and $MoS_2$ Raman peaks exhibit significant changes in their frequencies and linewidths inside the heterostructure. We attribute these changes to strain and charge transfer doping, as explained further below. The graphene *G* peak frequency ($\omega_G$) redshifts by 8-10 $cm^{-1}$ in the heterostructure, as seen in the 2D frequency map in Figure 1e. On the other hand, the *G'* peak blueshifts ($\omega_{G'}$) by 5-10 $cm^{-1}$ between $SiO_2$ and the heterostructure (Figure 1f). Both graphene peaks also exhibit significant broadening in the heterostructure compared to $SiO_2$, with the *G* peak broadening ($\Gamma_G$) from ~10 $cm^{-1}$ in bare graphene to ~25 $cm^{-1}$ in the heterostructure (Figure 1g). The linewidth of the *G'* peak $\Gamma_{G'}$, which is often correlated to graphene quality is ~25 $cm^{-1}$ in bare graphene and broadens to ~45 $cm^{-1}$ in the heterostructure. The Raman peaks of $MoS_2$ also undergo changes in intensity and frequency. Both *E*' (Figure 1i) and *A'* (Figure 1j) peaks blueshift in frequency by an average of 2 $cm^{-1}$, in the heterostructure. In addition, both peaks narrow by 2-3 $cm^{-1}$ in the heterostructure compared to bare $MoS_2$ (Figure 1k shows a 2D linewidth map of the *E*' peak).

Taken together, the maps in Figs. 1b-1k show that the Raman peaks of both graphene and $MoS_2$ shift in frequency when the two layers are in contact with each other. As-deposited graphene and $MoS_2$ on $SiO_2$ substrates are known to be p-type and n-type doped, respectively.[26-29] This charge state comes from interaction with the $SiO_2$ substrate and the ambient environment,[26, 27] and in the case of $MoS_2$ the additional electron density could also originate from sulfur vacancies.[28, 29] Thus when graphene is placed on to $MoS_2$, we expect $MoS_2$ to donate electrons to graphene, and that is what we infer from the data shown in Figure 1. Both the *G* and *G'* peaks in bare graphene are blueshifted in frequency (p-doped or hole-doped) compared to the frequency of pristine undoped graphene,[30] while they are redshifted within the heterostructure, suggestive of electron transfer from the underlying $MoS_2$. Similarly, the blueshift of the $MoS_2$ Raman peaks indicate electron transfer to the graphene.[31]

However, the linewidths of the graphene and $MoS_2$ Raman peaks do not follow the trends established for charge transfer doping. We observe broadening of the *G* peak by as much as 15 $cm^{-1}$, whereas it sharpens for both electron- and hole-doping due to the strong electron-phonon coupling of the *G* peak phonons. This coupling reduces the number of decay channels due to Pauli blocking.[30, 32] Considering that our graphene on $SiO_2$ is already p-doped, a slight peak narrowing is expected upon electron transfer from $MoS_2$ to graphene. However, the observation of significant broadening hints at a different mechanism, which we attribute to strain. In the case of $MoS_2$, upon hole-doping the Raman peaks have been observed to sharpen,[6, 33] in agreement with our observation of peak linewidth narrowing in the heterostructure (Fig. 1k).

## 4. Discussion

The Raman frequencies ($\omega$) are related to lattice strain ($\varepsilon$) by the Grüneisen parameter ($\gamma$), according to the following equation –

$$\Delta\omega = -2\gamma_i \omega_i^0 \varepsilon_i \tag{1}$$

$\gamma_i$ is the Grüneisen parameter corresponding to the frequency of peak $i$ ($\omega_i$), and $\omega_i^0$ is the frequency corresponding to zero strain. The Grüneisen parameter has been experimentally determined for both graphene[21, 34-37] and $MoS_2$.[38, 39] Average room temperature $\gamma$ values for uniaxial strain from the literature are 1.9, 2.6, 0.86, and 0.15 for the *G*, *G'*, *E'* and *A'* peaks respectively. Furthermore, the doping dependence of the graphene[30] and $MoS_2$[31] peak frequencies has been experimentally measured and found to be quasi-linear for electron and hole doping in $MoS_2$ and for hole doping in graphene,[30, 31] while it is non-linear for electron doping in graphene.[30] However, here we neglect electron doping in graphene because of many studies that have shown graphene to be p-doped on $SiO_2$ substrates.[40, 41] Assuming a linear relationship, we have the relation between the peak frequency and carrier concentration as follows-

$$\omega_i = k_n(i)n \tag{2}$$

Here $n$ is the carrier concentration and $k_n(i)$= -0.33 x $10^{13}$, -2.22 x $10^{13}$, -9.6 x $10^{13}$, -1 x $10^{13}$, for $i$= *E'*, *A'*, *G* and *G'* peaks, respectively. The constants $k_n(i)$ have been determined empirically in Refs. 30 and 31. Since the peak frequencies depend on both strain and carrier concentration, we combine equations (1) and (2) to obtain a pair of equations:

$$\Delta\omega_1 = -2\gamma_1\omega_1^0\varepsilon + k_n(1)n \qquad (3)$$

$$\Delta\omega_2 = -2\gamma_2\omega_2^0\varepsilon + k_n(2)n \qquad (4)$$

Here the subscripts 1 and 2 refer to the pair of peaks *G*, and *G'* for graphene, and *E'* and *A'* for $MoS_2$, respectively. Equations (3) and (4) can be rearranged and expressed in terms of $\varepsilon$ and *n* as follows –

$$\varepsilon = \frac{k_n(1)\Delta\omega_2 - k_n(2)\Delta\omega_1}{2\gamma_1\omega_1^0 k_n(2) - 2\gamma_2\omega_2^0 k_n(1)} \qquad (5)$$

$$n = \frac{\gamma_1\omega_1^0\Delta\omega_2 - \gamma_2\omega_2^0\Delta\omega_1}{\gamma_1\omega_1^0 k_n(2) - \gamma_2\omega_2^0 k_n(1)} \qquad (6)$$

Equations (5) and (6) enable the calculation of Raman peak frequencies for constant strain and carrier concentration. In order to perform this calculation, one must first establish the Raman peak frequencies for unstrained and undoped graphene and $MoS_2$. For this we choose the Raman frequencies from suspended graphene[19, 42-44] and $MoS_2$,[45-47] which can be considered as their pristine unperturbed state. A plot between the calculated graphene *G'* and *G* peak frequencies, and between the $MoS_2$ *E'* and *A'* peaks can then be drawn with superimposed constant strain ($\varepsilon$) and constant carrier concentration (*n*) axes as shown in Figure 2. The calculated constant $\varepsilon$ and constant *n* axes are drawn on the $\varepsilon$-*n* plot for increments of 0.1% and $0.5 \times 10^{13}$ cm$^{-2}$, respectively, with positive (negative) strain values corresponding to tension (compression). The straight lines denote zero strain and doping, and the green square data point corresponds to the Raman peak frequency for suspended graphene and $MoS_2$. We now plot the experimentally measured Raman peak frequencies (Figs. 1 and S1) from graphene and $MoS_2$ from both the vertical heterostructures on the $\varepsilon$-$n$ plot. (Figure 3).

Comparing the data between bare graphene (blue circles for Het-1 and blue triangles for Het-2) and graphene on $MoS_2$ (red circles for Het-1 and red triangles for Het-2) points in Figure 3a, one immediately sees that they are different and shifted from each other. The bare graphene in both Het-1 and Het-2 is p-doped, with an average hole concentration of ~1.4 x $10^{13}$ $cm^{-2}$. In addition, bare graphene is also under tensile strain (up to 0.3%), with the average strain slightly higher in Het-2 than Het-1. For the graphene in the heterostructure, the *G* and *G'* peaks are redshifted and blueshifted in frequency, respectively, implying a reduction in hole concentration from ~1.4 x $10^{13}$ $cm^{-2}$ to an average of 0.4 x $10^{13}$ $cm^{-2}$. Furthermore, the tensile strain in the bare graphene converts to compressive strain in both heterostructures. The black arrow in Figure 3a indicates the overall shift of the Raman peaks from bare graphene to graphene in the heterostructures.

Figure 3b shows the $\varepsilon$-$n$ plot with the $MoS_2$ *E'* and *A'* Raman peaks from Het-1 and Het-2. Unlike graphene, where the peak frequencies in both heterostructures are similar, bare $MoS_2$ in the two heterostructures exhibits differences in charge and strain states. However, within both heterostructures the $MoS_2$ donates electrons to graphene and undergoes compression to varying degrees. Bare $MoS_2$ in Het-1 (blue circles in Figure 3b) is n-doped with an average electron density of ~1 x $10^{13}$ $cm^{-2}$ and is under tensile strain (average ~0.38%). It undergoes compression (strain relief) in the heterostructure to an almost unstrained state. The $MoS_2$ in Het-1 donates electrons to graphene and its electron density is lowered from ~1 x $10^{13}$ in bare $MoS_2$ to ~0.5 x $10^{13}$ $cm^{-2}$ in the heterostructure. Bare $MoS_2$ in Het-2 is slightly p-doped and under tension and undergoes slight compression and p-doping within the heterostructure.

The Raman frequencies, strains and carrier concentrations in bare graphene and $MoS_2$ and within the two heterostructures are tabulated in Table 1. Graphene in both the heterostructures behaves in a similar fashion with respect to strain and charge transfer (undergoes compression and gains electrons from $MoS_2$), and $MoS_2$ loses electrons to graphene while also undergoing compression in the heterostructure. While the graphene Raman peak frequencies are very similar in the two heterostructures, the $MoS_2$ in Het-2 is influenced more by doping (data points shift towards the right in Fig. 3b) than in Het-1 where the influence of strain is greater than doping (data points shift upwards in Fig. 3b).

Although it is not surprising to find that the electron-rich $MoS_2$ donates electrons to electron-poor graphene, the observation of compressive strain in both layers (up to 0.3%) in two different heterostructures is noteworthy. Graphene has a negative Coefficient of Thermal Expansion (CTE) of ~ $-8 \times 10^{-6}$ $K^{-1}$ at 300 K,[48, 49] while the CTE for $MoS_2$ is positive (~$6 \times 10^{-6}$ $K^{-1}$ at 300K).[49] The CTEs of both graphene and $MoS_2$ are an order of magnitude larger than that of $SiO_2$ (~$0.6 \times 10^{-6}$ $K^{-1}$).[50] Since our samples undergo thermal annealing (350 ºC) after mechanical exfoliation, one could expect the emergence of strain in the graphene and $MoS_2$ owing to the differences between their CTEs and the much smaller CTE of the $SiO_2$ substrate, as well as the differences between the lattice constants of graphene (2.40 Å) and $MoS_2$ (3.12 Å). Indeed, large tensile and compressive strains (up to 0.7%) have been observed in bare $MoS_2$ and graphene on substrates upon heating and cooling.[20, 21, 51] Note that compressive and tensile strains are also observed in as-transferred graphene and $MoS_2$ without any thermal annealing.[15, 18] Our observed strain values in bare graphene and $MoS_2$ are within the range of strains published in the literature, and also match the calculated values (~0.1%) for cooling from 350 ºC to room temperature.

It is important to reiterate that the strains and carrier concentrations in our graphene/$MoS_2$ heterostructures are calculated based on the Raman frequencies of the respective suspended layers, which we assume to be unstrained and undoped. The absolute values of the initial charge state and strain cannot be known unless they are measured independently. However, we can still compare the relative increases or decreases of Raman frequencies in the heterostructures compared to bare $SiO_2$, and therefore our observation of electron transfer from $MoS_2$ to graphene and compressive strain in the heterostructures is valid. While these observations suggest a universal trend, the magnitudes of charge transfer and strain are evidently different between Het-1 and Het-2.

So far, we have discussed the differences in Raman peak frequencies inside and outside the heterostructure. For $MoS_2$ we can also analyze the PL emission intensities and energies. As shown in Figures 1l and 1m, the PL emission gets quenched drastically by over 90% from bare $MoS_2$ to the heterostructure. Quenching of the $MoS_2$ PL intensity has been observed before in graphene/$MoS_2$ heterostructures and is generally attributed to the coupling between the two layers.[3, 52-54] The PL emission in $MoS_2$ occurs from a recombination of photo-excited excitons corresponding to an optical bandgap of ~1.83 eV.[55, 56] The PL emission band typically also contains a second peak at a lower energy, corresponding to a recombination of trions (excitons formed by two electrons/one hole or two holes/one electron).[57] As deposited $MoS_2$ on $SiO_2$ is typically electron-rich (n-type), resulting in an abundance of negative trions, hence one expects to see a higher trion peak intensity in the PL emission. Moreover, in general the PL emission energy redshifts (blueshifts) with hole (electron) doping. But the electronic structure of a material also influenced by strain. The bandgap of monolayer $MoS_2$ depends approximately linearly on the applied strain at rates ranging from -25 to -45 meV/%.[10, 13, 14] Tensile (compressive) strain typically results in a redshift (blueshift) of the emission energy, along with broadening of the PL emission peaks.

In Het-1 we observe a blueshift of the emission by up to 20 meV as shown in the 2D PL emission map in Figure 4a. By extracting a couple of representative spectra from the 2D PL emission map (Fig. 4a) and by performing Lorentzian lineshape analysis, we can see blueshifts in both the $A^-$ (trion) and A (neutral exciton) peak energies (Fig. 4b), suggesting that the PL from the $MoS_2$ in Het-1 is influenced more by strain than by doping. Furthermore, the magnitude of the blueshift (ranging from 10 to 20 meV) corresponds to strains ranging from 0.2 to 0.5%; this matches very well with the observed Raman frequencies. In contrast, the Raman spectra from the $MoS_2$ in Het-2 show that it is influenced more by doping than by strain (Fig. 3b). This can also be seen in the PL emission maps in Fig. 4c, where we plot the 2D emission maps for both *A* and *B* excitons. Note that the broad PL emission peaks in Het-2 make it difficult to deconvolve the trion and neutral exciton peaks, hence we fit it to a single Lorentzian peak labeled *A*. Fits to representative spectra from Het-2 reveal a slight redshift, which we attribute to hole doping.

To see if it is possible to modulate the carrier concentration or the strain in the heterostructures, we investigate the effect of additional thermal annealing. Figure 5 shows the Raman peak frequencies of graphene and $MoS_2$ in Het-1 (Figures 5a and 5b) and Het-2 (Figures 5c and 5d) after thermal annealing at 350 °C for 30 minutes under flowing argon at atmospheric pressure. The Raman peak frequencies corresponding to the un-annealed (same as in Figure 3) and annealed states are shown as light and dark data points, respectively. In both heterostructures we see that bare graphene becomes more p-doped, with similar increases in hole concentration (from ~1.4 to ~1.6 x $10^{13}$ $cm^{-2}$ on average) upon thermal annealing (Figures 5a and 5c). Although both heterostructures are annealed under flowing argon gas, the process is performed at atmospheric pressure and the increase in hole concentration can be attributed to doping from residual oxygen in the chamber.[16, 18] Interestingly, however, annealing affects the strain differently in bare graphene. After the additional annealing, the tensile strain in bare graphene in Het-1 increases slightly, while it decreases to an almost unstrained state in Het-2. Within the heterostructure, the additional annealing causes the graphene to undergo compressive strain and become less hole-doped by accepting electrons from the underlying $MoS_2$ layer. The overall direction of change in graphene peak frequencies is the same as that observed prior to thermal annealing (data points move to the left and upwards on the $\varepsilon$-$n$ plot, i.e. compression and reduction in hole concentration between bare graphene and graphene in the heterostructure). But the magnitude of strain is different in the two heterostructures - graphene in Het-2 undergoes much more compression (up to 0.5%) after the additional annealing compared to Het-1.

In the case of $MoS_2$, thermal annealing causes both heterostructures to behave in a similar fashion, with bare $MoS_2$ undergoing compression and p-doping. However, the additional annealing affects bare $MoS_2$ in Het-1 much more than in Het-2. It undergoes significant p-doping (up to $2 \times 10^{13}$ cm$^{-2}$) and compression (0.4% on average) to an almost unstrained state. Within the heterostructure the $MoS_2$ in Het-1 undergoes further compressive strain (~0.1% on average) and hole-doping. The annealing does not appear to affect the $MoS_2$ in Het-2 as much as in Het-1, although the general trend is the same (p-doping and compression). After thermal annealing the PL emission from $MoS_2$ in both heterostructures remains blueshifted compared to bare $MoS_2$.

The $\varepsilon$-$n$ plots in Figure 5 demonstrate an approach to modulate the strain and carrier concentration in graphene/$MoS_2$ heterostructures through thermal annealing. While we have only shown the effects of thermal annealing under argon at atmospheric pressure, other forms of annealing, for example in vacuum, photonic annealing, or exposure to other environments could be employed to tune the strain and carrier concentration in graphene/$MoS_2$ and other 2D heterostructures. Figure 5 also draws attention to an important issue regarding sample preparation – reproducibility. Our results show that samples that have been prepared under similar conditions (mechanical exfoliation followed by transfer) behave quite differently depending upon thermal annealing. The reason for this difference is unclear and requires further experimental and theoretical investigations.

Finally, the variations in the carrier concentrations and strain states in the heterostructures suggest that the discrepancies reported in the literature could be attributed to different charge and strain states of graphene and $MoS_2$ after sample preparation. Moreover, the clear differences between two heterostructures that were prepared by the same method highlight inherent variabilities in the mechanical transfer processes and point to the need for a fast and convenient method to characterize these differences. Our usage of $\varepsilon$-$n$ plots shows that it is important to concurrently measure strain and doping, both of which can easily affect atomically thin materials. Although here we considered uniaxial strains in the 2D layers, we cannot exclude biaxial strain and further experimental and theoretical analysis could provide more insights. The van der Waals interface between 2D materials is highly sensitive to the processing conditions, and care must be taken during sample preparation and characterization. The method described herein can also be extended to other materials towards strain engineering and tuning the carrier concentration in 2D heterostructures.

## Conclusions

We have studied coupling-induced charge transfer doping and strain in two graphene/$MoS_2$ heterostructures using Raman and PL spectroscopy. By using correlation analysis between the *G'* and *G* peaks in graphene and between the *E'* and *A'* Raman peaks in $MoS_2$, we show that graphene ($MoS_2$) is p-doped (n-doped) on $SiO_2$ and that there is a transfer of electrons from $MoS_2$ to graphene. Moreover, we find bare graphene and $MoS_2$ on $SiO_2$ to be affected varying degrees of strains, likely induced during the transfer process. Within the heterostructure we find both graphene and $MoS_2$ to undergo compressive strain. Interestingly, both charge concentration and strain can be tuned to a certain degree with thermal annealing. Our study shows that it is important to consider both mechanical and electronic coupling when characterizing van der Waals interfaces in 2D heterostructures, and at the same time it demonstrates a way to modulate strain and carrier concentration through annealing.

## Acknowledgements

We gratefully acknowledge funding from the Air Force Office of Scientific Research under LRIR No. 16RXCOR322.

## Corresponding Authors

*Rahul Rao: rahul.rao.1.ctr@us.af.mil, Jyoti Katoch: jkatoch@andrew.cmu.edu

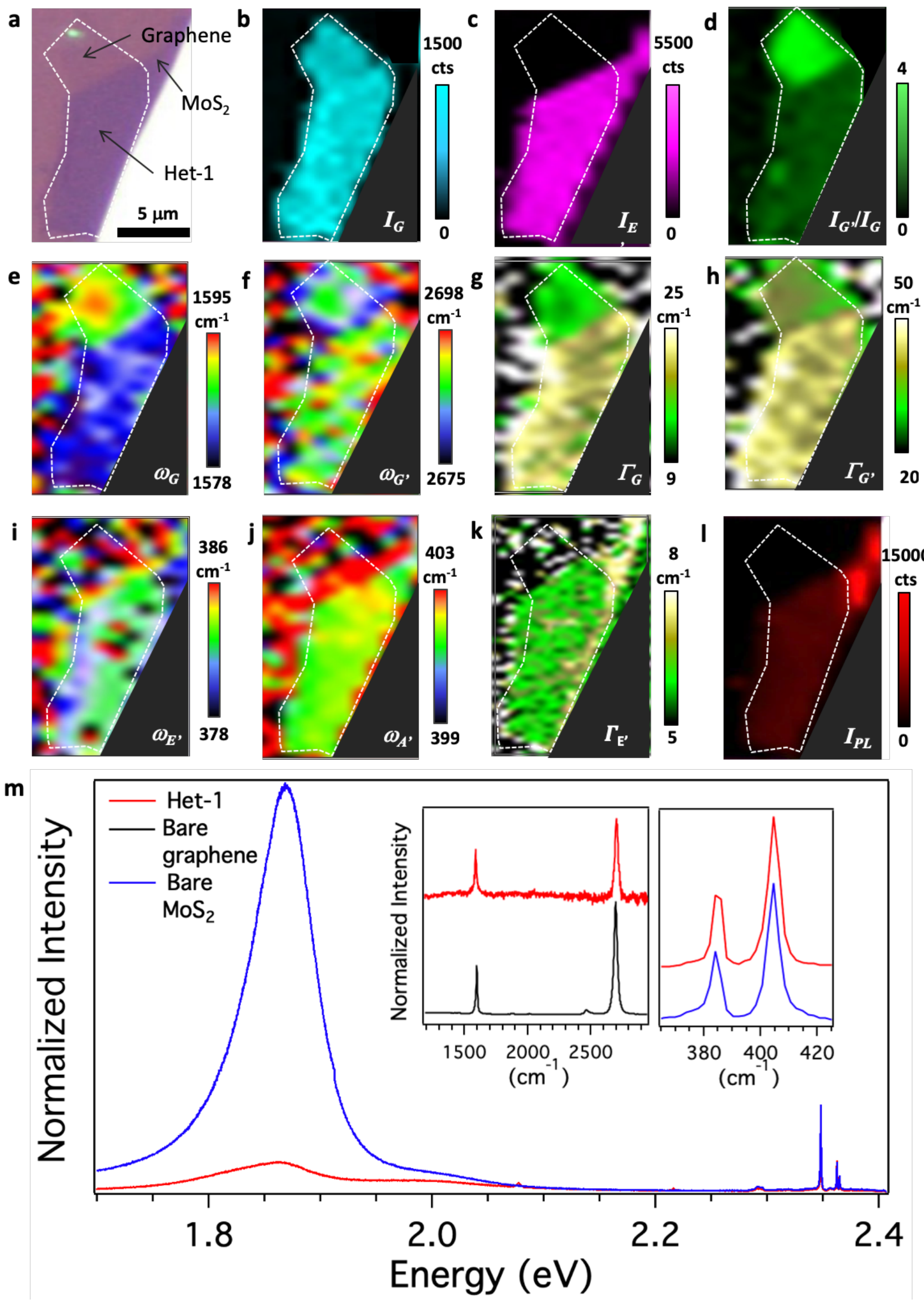

Figure 1. Optical characterization of a graphene/$MoS_2$ heterostructure on $SiO_2$ (Het-1). (a) Optical microscope image of Het-1. The top layer (monolayer graphene) is shown by the dashed outline. 2D Raman intensity maps are shown for the (b) graphene *G* peak and (c) $MoS_2$ *E'* peaks. (d) Intensity ratio of the graphene *G'* and *G* peaks showing a suppression of the *G'* peak intensity within the heterostructure. (e), (f) Maps showing frequencies of the graphene *G* and *G'* peaks, respectively. (g), (h) Maps showing linewidths (FWHM) of the graphene *G* and *G'* peaks, respectively. (i), (j) Maps showing frequencies of the $MoS_2$ *E'* and *A'* peaks, respectively. (k), Map

showing linewidths (FWHM) of the $MoS_2$ *E'* peak. (l) Map of the PL emission intensity from the $MoS_2$. (m) Representative PL emission spectra (normalized to the $SiO_2$ substrate Raman peak intensity) from $MoS_2$ on $SiO_2$ (top trace) and the graphene/$MoS_2$ heterostructure (bottom trace). The left inset shows the Raman peaks from the bare graphene (bottom trace) and graphene in Het-1 (top trace). The right inset shows the Raman peaks from the bare $MoS_2$ (bottom trace) and $MoS_2$ within Het-1 (top trace).

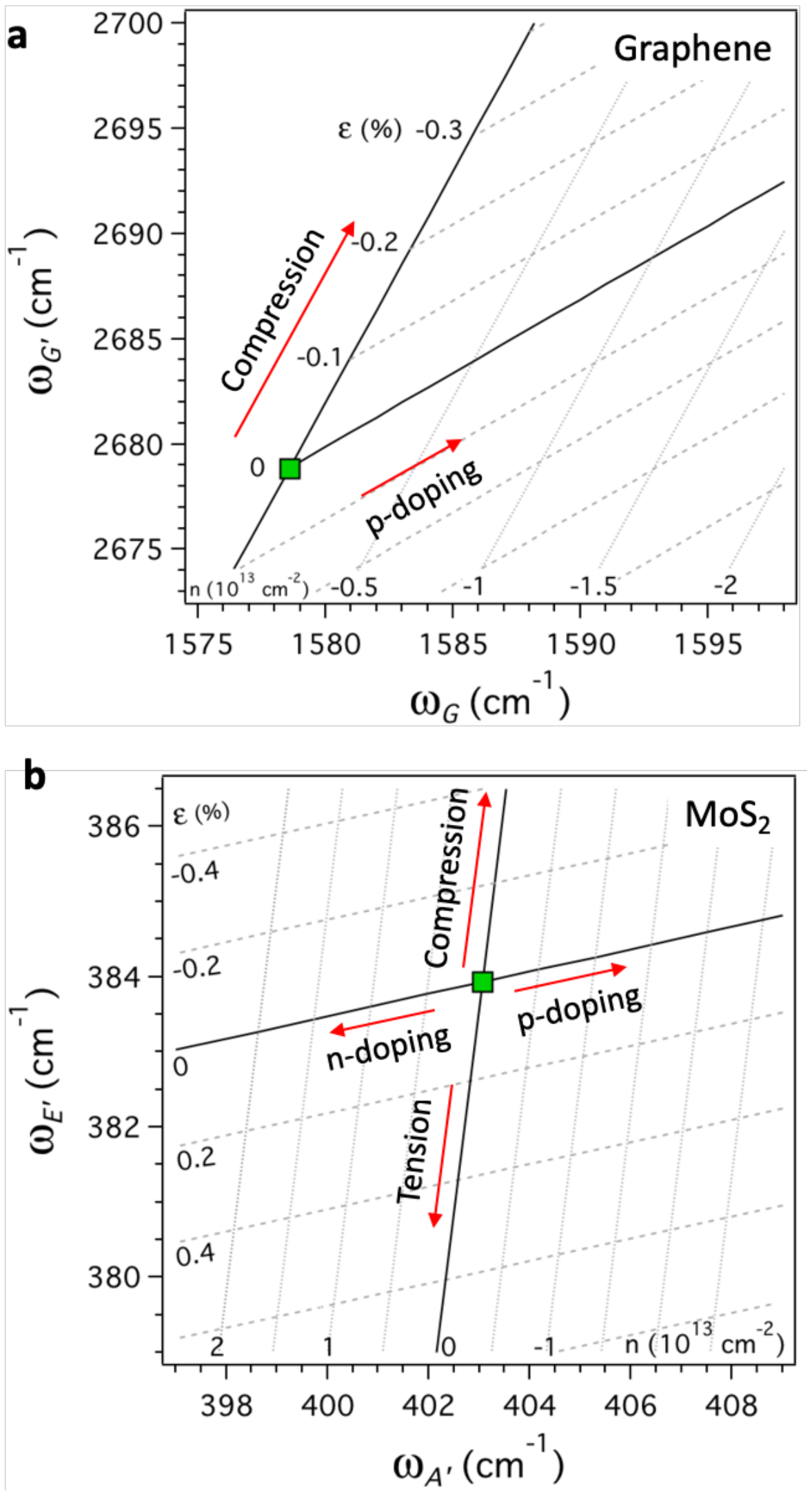

Figure 2. *ε*-*n* plots for (a) graphene and (b) $MoS_2$ calculated using equations (5) and (6) for various strain and carrier concentration values. The green square data points correspond to undoped and unstrained graphene and $MoS_2$, which we assume to be the Raman peak frequencies from suspended layers.

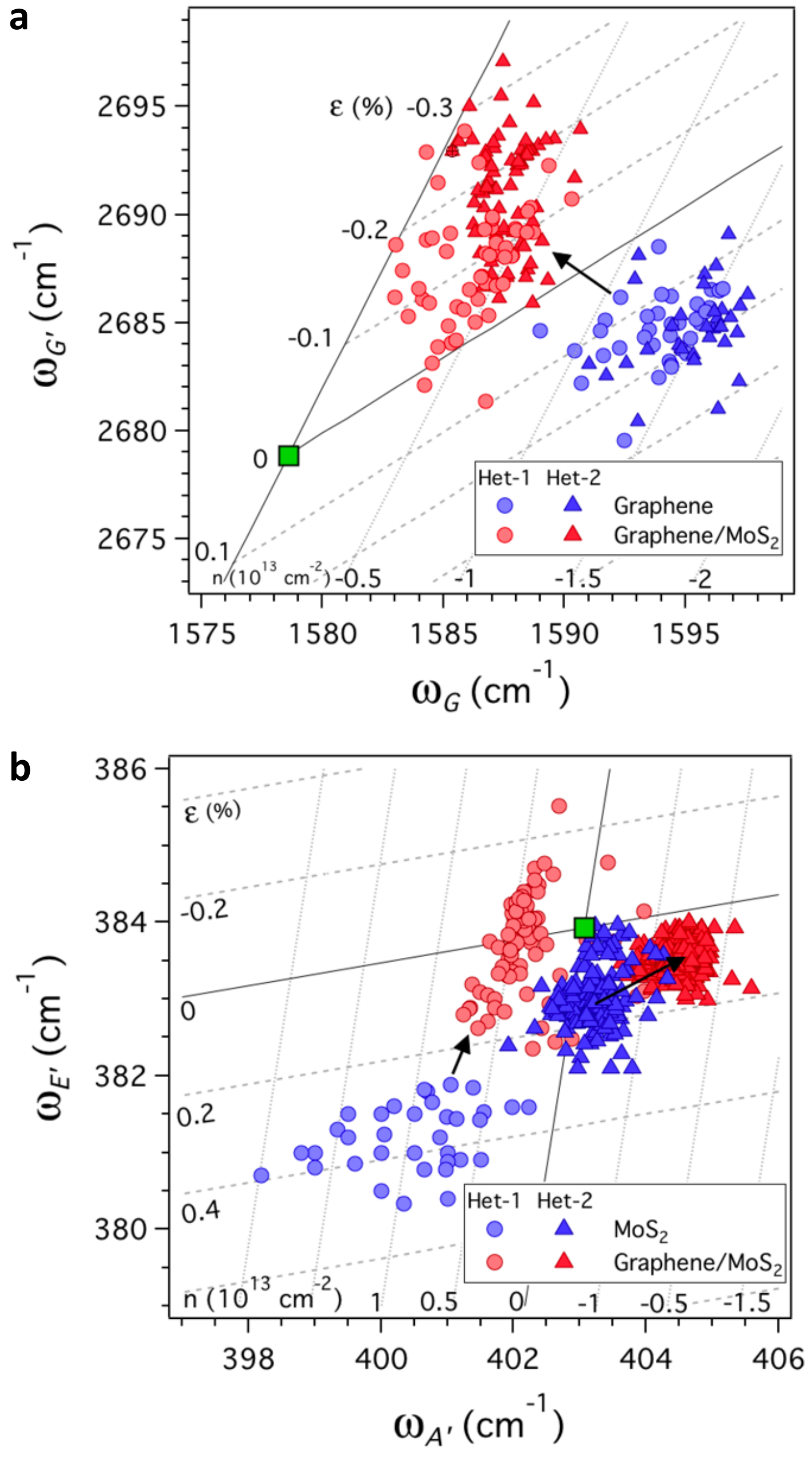

Figure 3. Strain-doping (*ε*-*n*) plots for two different graphene/$MoS_2$ heterostructures (Het-1, circles and Het-2, triangles) obtained by plotting frequencies of the (a) *G’* vs. *G* peak for graphene on $SiO_2$ and graphene in the heterostructure, and (b) *A’* vs. *E’* peak for $MoS_2$ on $SiO_2$ and $MoS_2$ in the heterostructure. The green data points in (a) and (b) correspond to Raman frequencies of suspended graphene and $MoS_2$, respectively. The shift in frequencies from bare $SiO_2$ to the heterostructure are indicated by the arrows.

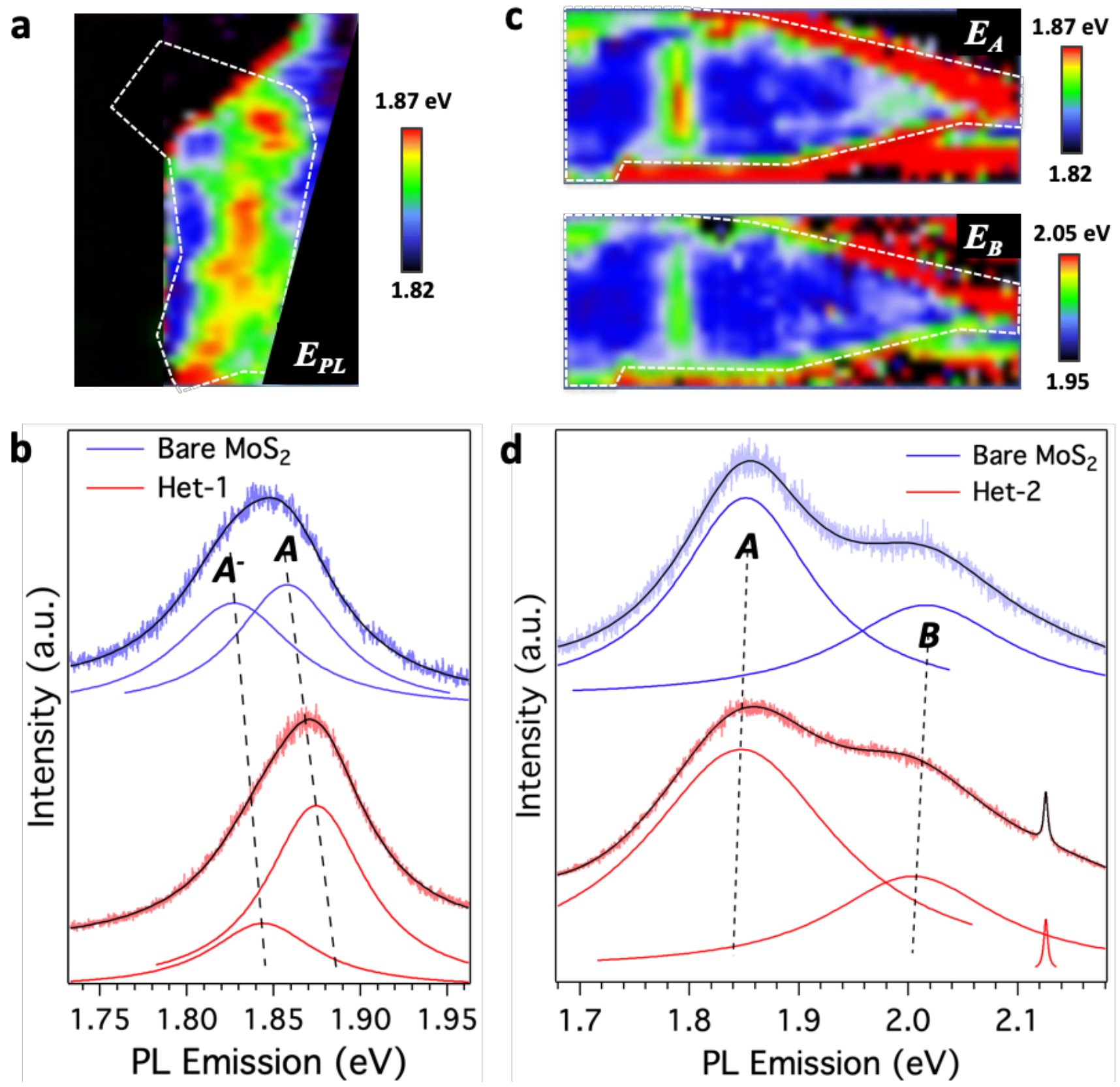

Figure 4. 2D PL emission maps and representative Raman spectra from (a) Het-1 and (b) Het-2. The overall blueshift and redshifts observed in Het-1 and Het-2 indicate greater influences from compressive strain and doping, respectively.

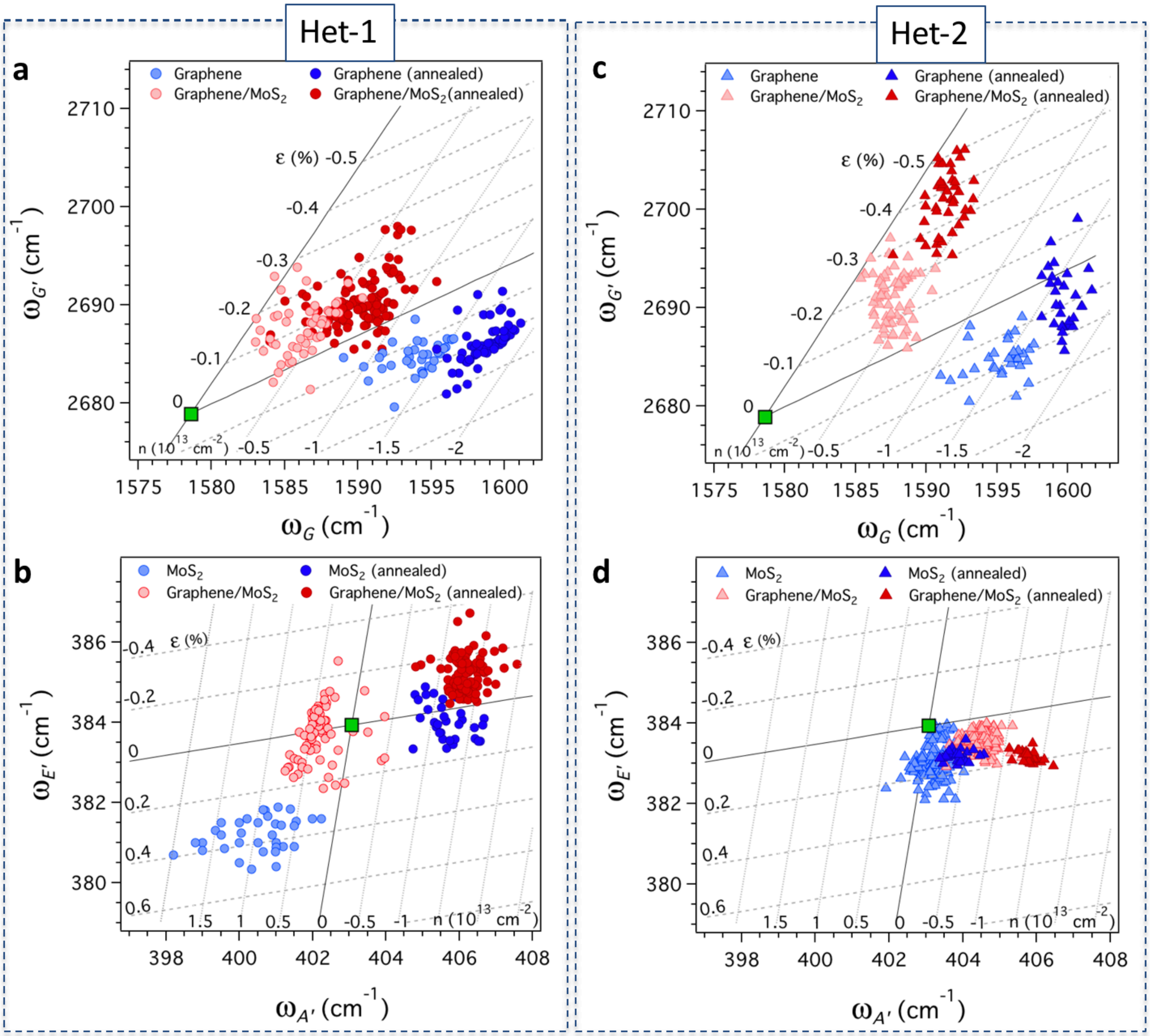

Figure 5. Raman peak frequencies from Het-1 (a, b) and Het-2 (c, d) after thermal annealing at 350 °C under atmospheric pressure argon. The lighter (darker) data points correspond to before (after) annealing. The *ε*-*n* plots are drawn on the same scale to show how thermal annealing affects the strain and carrier concentration in graphene/$MoS_2$ heterostructures prepared under similar conditions.

**Table 1**. Range of Raman peak frequencies and their average values (in bold, within parentheses) measured from bare graphene (top) and $MoS_2$ (bottom) and within the two heterostructures. Also included are the estimated ranges of strain and carrier concentration. Positive (negative) numbers correspond to tension and electron- (compression and hole-) doping.

| | | | | | |
|---|---|---|---|---|---|
| **Graphene** | | | | | |
| | | *G* ($cm^{-1}$) | *G'* ($cm^{-1}$) | Strain (%) | Carrier concentration (x $10^{13}$ $cm^{-2}$) |
| Het-1 | Bare | 1587 to 1596 (***1594.3***) | 2680 to 2688 (***2684.7***) | 0.03 to 0.28 (***0.12***) Compression ↓ | -0.7 to -1.5 (-***1.3***) Gains $e^-$ ↓ |
| | Heterostructure | 1583 to 1590 (***1586.3***) | 2686 to 2694 (***2687.7***) | 0.1 to -0.28 (***-0.15***) | 0 to -0.5 (-***0.4***) |
| Het-2 | Bare | 1591 to 1597 (***1595.3***) | 2683 to 2686 (***2684.7***) | 0.02 to 0.3 (***0.15***) Compression ↓ | -1.1 to -1.7 (-***1.5***) Gains $e^-$ ↓ |
| | Heterostructure | 1585 to 1590 (***1587.7***) | 2686 to 2695 (***2691***) | 0 to 0.33 (***-0.18***) | 0 to -0.6 (-***0.3***) |
| **$MoS_2$** | | | | | |
| | | *E'* ($cm^{-1}$) | *A'* ($cm^{-1}$) | Strain (%) | Carrier concentration (x $10^{13}$ $cm^{-2}$) |
| Het-1 | Bare | 380 to 381.5 (***381***) | 398 to 402 (***400***) | 0.26 to 0.5 (***0.4***) Compression ↓ | 0 to 2 (***1***) Loses $e^-$ ↓ |
| | Heterostructure | 382 to 385 (***383***) | 400.5 to 402 (***401.5***) | 0.2 to -0.2 (***0***) | -1 to 0.5 (***0.4***) |
| Het-2 | Bare | 381.5 to 383.5 (***382.5***) | 402 to 404 (***401***) | 0 to 0.3 (***0.15***) Compression ↓ | 0.4 to -1 (***0.3***) Loses $e^-$ ↓ |
| | Heterostructure | 382.5 to 383.5 (***383***) | 402.5 to 405 (***404***) | 0 to 0.2 (***0.1***) | -0.5 to -1.2 (***0.1***) |